\documentclass{optica-article}

\journal{opticajournal} 

\articletype{Research Article}

\usepackage{lineno}
\usepackage{subfigure}
\usepackage{svg}
\usepackage{textcomp}

\newcommand{\SUm}[1]{mPa/$\sqrt{\textrm{Hz}}$}
\newcommand{\SUu}[1]{\textmu Pa/$\sqrt{\textrm{Hz}}$}

\begin{document}

\title{
Fibre-coupled photonic crystal hydrophone
}

\author{Lauren R. McQueen,\authormark{1,2,3*}, Nathaniel Bawden,\authormark{1,2,3}, Benjamin J. Carey,\authormark{1,2,3} Igor Marinkovi\'{c},\authormark{1,2,3}, Warwick P. Bowen,\authormark{1,2,3\textsuperscript\textdagger}, Glen I. Harris,\authormark{1,2}.}

\address{\authormark{1}School of Mathematics and Physics, The University of Queensland, St. Lucia, Queensland 4072, Australia \\
\authormark{2}Australian Research Council Centre of Excellence for Engineered Quantum Systems, The University of Queensland, St. Lucia, Queensland 4072, Australia\\
\authormark{3}Australian Research Council Centre of Excellence in Quantum Biotechnology, The University of Queensland, St. Lucia, Queensland 4072, Australia}

\email{\authormark{*}l.mcqueen@student.uq.edu.au} 
\email{\authormark{\textdagger}w.bowen@uq.edu.au}

\begin{abstract*}

Many applications, including medical diagnostics, sonar and navigation rely on the detection of acoustic waves. Photonic hydrophones demonstrate comparable sensitivity to piezoelectric-based hydrophones, but with significantly reduced size, weight and power requirements. In this paper we demonstrate a micron-sized free-standing silicon photonic hydrophone. We demonstrate sensitivity on the order of $\sim$\SUm{1} from 10-200~kHz, with a minimum detectable pressure of 145~\SUu{1} at 22~kHz. We also deployed our hydrophone in a wave flume to evaluate its suitability for underwater measurement and communication. Our hydrophone matches the sensitivity of commercial hydrophones, but is many orders of magnitude smaller in volume, which could enable high spatial resolution imaging of micron-sized acoustic features (\textit{i.e.}, living cell vibrations). Our hydrophone could also be used in underwater communication and imaging applications.

\end{abstract*}




\section{Introduction}\label{Section: introduction} 

The detection of acoustic waves is used in a range of applications, including medical diagnostics \cite{wissmeyer2018looking}, industrial processes, such as quality control and defect assessment for construction \cite{fischer2016optical}, sonar, navigation \cite{kapoor2018acoustic} and trace gas sensing \cite{wu2017beat}. Most applications require high directionality, as well as high spatial and temporal resolution. This has motivated the development of acoustic sensors that operate at ultrasonic frequencies, corresponding to short acoustic wavelengths, and micro-scale sensing devices that can resolve waves close to or beyond their diffraction limit \cite{miller2016ultrasound}.  

Many hydrophones used in commercial applications are based on a piezoelectric-transducer mechanism \cite{pauer1974flexible,shi2020design,hurrell2012piezoelectric,kaya1989acoustic}. However, degradation in acoustic sensitivity becomes a significant obstacle when operating at high frequencies and with devices with smaller sensing areas \cite{ballantine1996acoustic}. This has motivated progress in photonic hydrophones \cite{preisser2016all,guggenheim2017ultrasensitive,li2021cavity,basiri2019precision}. 
The main advantage of photonic hydrophones is that they significantly reduce size, weight and power (SWaP) requirements, while demonstrating comparable sensitivity to piezoelectric-based hydrophones. This is advantageous in scenarios that require the deployment of hydrophones on drones for surveillance, or arrays of many hydrophones for underwater communication and sonar, which could also enable high spatio-temporal resolution. The reduced size requirements can also be advantageous for biomedical applications. 

Photonic hydrophones have been widely explored in different regimes. Fibre-optic hydrophones have been explored since the 1970s \cite{kraakenes1989sagnac,lim1999fiber,kipergil2023fiber}, and by the 1990s, they had been developed to the point where they had outstanding sensitivity compared to piezoelectric-based hydrophones \cite{kirkendall2004overview,kirkendall2007distributed}. However, to achieve this they require tens to hundreds of metres of fibre interaction length, which renders them relatively bulky and limited in bandwidth. To overcome both these limitations, fibre-laser hydrophones were developed, which achieved comparable sensitivity with less than a centimetre of fibre interaction length \cite{foster2005fiber,cranch2009fiber}. However, mechanical constraints in mounting and managing the distributed feedback structure, which these hydrophones are based on, limited the bandwidth to about 10~kHz. Then, with the evolution of silicon-on-insulator (SOI) technology, hydrophones have been designed to have all the mechanical components on chip, which has advanced the performance (improved sensitivity and increased bandwidth) of photonic hydrophones by enabling the fabrication of micro-scale silicon resonator-based devices with nano-scale features that can be mass-fabricated on-chip \cite{celler2003frontiers}. SOI hydrophones typically operate in the MHz frequency range and demonstrate $\sim$\SUm{1} sensitivity \cite{guggenheim2017ultrasensitive,monifi2013ultrasound,shnaiderman2020submicrometre,westerveld2021sensitive,nagli2023silicon}, while air-coupled optomechanical acoustic sensors have demonstrated $\sim$\SUu{1} sensitivity \cite{basiri2019precision} which is boosted by the narrow-band mechanical resonance \cite{forstner2012cavity,yan2023force,pruessner2018optomechanical}. SOI-fabricated photonic hydrophones are also easily-scalable, and have been demonstrated for sonar \cite{westerveld2021sensitive} and  photoacoustic imaging applications \cite{sabuncu2023ultrafast,li2019disposable,hazan2022silicon}. The smallest optical cavity is a single defect cavity in a photonic crystal (PhC) \cite{nair2010photonic}, however PhCs are yet to be integrated into hydrophones with entirely on-chip mechanics. 

\begin{figure}[htbp]
\centering\includegraphics[width=\textwidth]{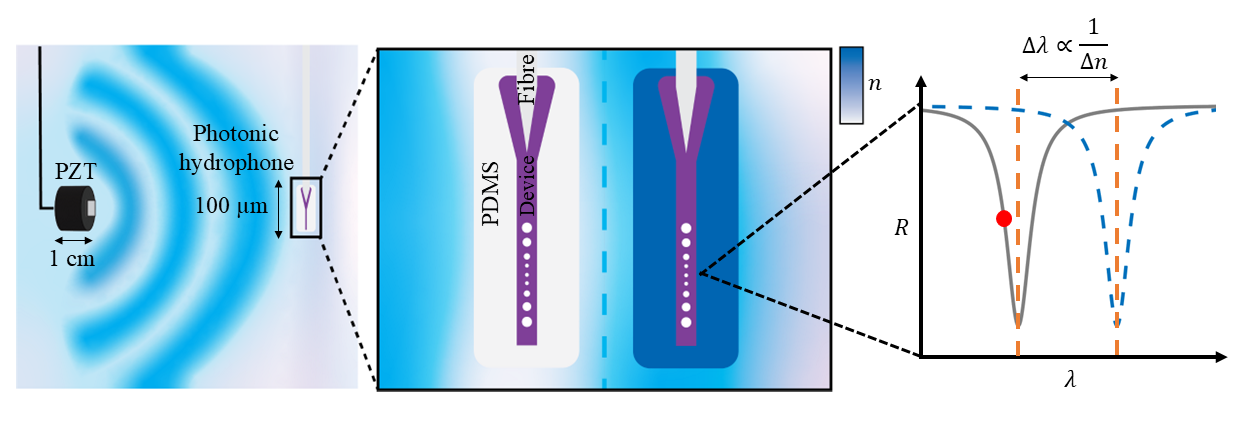}
\caption{\label{Fig: sensor schematic} Schematic of photonic hydrophone response to an acoustic signal. A piezoelectric transducer (PZT) and our hydrophone are submerged underwater. The PZT transduces an electrical signal into an acoustic signal, which when incident on the hydrophone, causes a change in the refractive index of the PDMS that encapsulates the silicon device. This induces a shift in the resonant wavelength of the defect region of the silicon device, which can be monitored via optical read-out.}
\end{figure}

In this paper we present the development of a micron-sized free-standing 1D PhC-based silicon photonic hydrophone. Our nano-fabricated sensor is composed of a 1D PhC with a defect cavity, with a tapered optical fibre that couples light into the cavity. The entire sensor with the glued fibre is encapsulated in polydimethylsiloxane (PDMS), a soft polymer which facilitates the sensor response to acoustic pressure (see Fig. \ref{Fig: sensor schematic}). We characterise the response of our photonic hydrophone from 10-200~kHz, which extends from the 10~Hz-100~kHz range studied with other PhC-based hydrophones \cite{lorenzo2021optical,kilic2011miniature,wong2018design}, and achieve a minimal detectable pressure of 145~\SUu{1}. Our hydrophone is the first with an entirely fibre-integrated PhC and consequently, the sensing area is 10$^4$ times smaller than previously demonstrated PhC-based hydrophones \cite{lorenzo2021optical}. 

Our photonic hydrophone is many orders of magnitude smaller in volume than piezoelectric-based hydrophones yet demonstrates comparable sensitivity. This could enable non-destructive high-resolution imaging of sub-micron sized acoustic features, such as living cell vibrations \cite{pelling2004local,jelinek2008measurement,sun2022encapsulated,sun2023whispering}, defects in the semiconductor micro-structures \cite{ensminger2024ultrasonics}, and inspecting integrated circuit packaging \cite{aryan2018overview}. Our hydrophone also significantly reduces SWaP requirements, which presents it as an alternative to piezoelectric-based hydrophones for underwater communication and imaging applications \cite{shi2020design,kaya1989acoustic}. We deploy our hydrophone in a wave flume to evaluate its suitability for underwater measurement and communication, and evaluate the bandwidth and signal decay rate of our hydrophone.

\section{Modelling of Hydrophone Response to Acoustic Signal}\label{Section: modelling}

\subsection{Hydrophone Architecture, Fabrication and Preparation}

\begin{figure}[htbp]
\centering\includegraphics[width=0.9\textwidth]{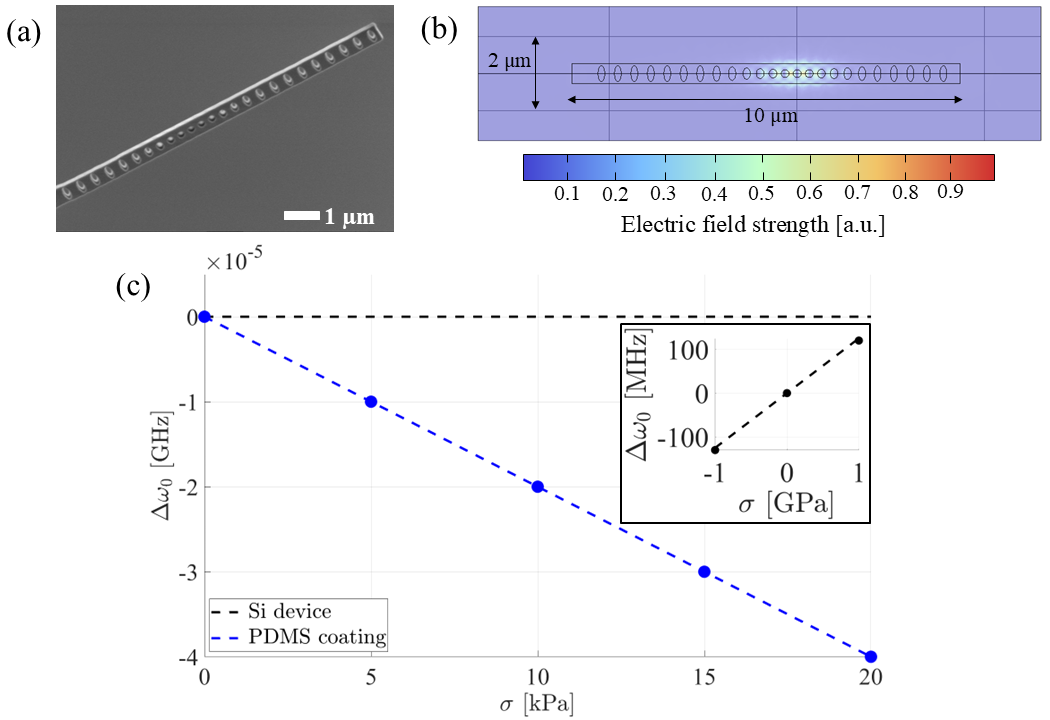}
\caption{\label{Fig: modelling figure} (a) SEM of PhC region of photonic hydrophone. (b) Electric field inside and around PhC region of device and coating.  (c) Plot of the change in eigenfrequency of the optical mode of our hydrophone $\omega$ against an applied pressure $\sigma$, for the photoelastic contributions from both the silicon device (black) and the PDMS coating (blue). These results are obtained from the FEM simulation. The inset figure displays the contribution from the silicon device to the eigenfrequency shift on GPa-scale.}
\end{figure}

The device architecture and fabrication procedure is very similar to that demonstrated in \cite{chan2012optimized,marinkovic2021hybrid}. The device includes a paddle which a tapered optical fibre is bonded to, a waveguide which extends out from the paddle and which the tapered fibre optically couples to, and a 1D PhC pattern etched at the end of the waveguide (see Fig.~\ref{Fig: modelling figure}(a)). The size of the etched holes decreases towards the centre, forming the defect region, and the larger holes on either side act as mirrors. This forms an optical cavity, whose resonant wavelength is sensitive to changes in the refractive index of the surrounding material \cite{chan2012optimized}. 

The device fabrication process involves using electron beam lithography (EBL) to define the device pattern on the SOI wafer (consisting of a layer of silicon separated from the bulk material by a layer of SiO$_2$), using electron beam resist as a mask for reactive ion etching (RIE) of the silicon. Since the device is connected to the rest of the chip with small ($\sim$100 nm) tethers, vapour phase hydroflouric (HF) etching is used to etch away the SiO$_2$ to suspend the devices from the chip without breaking them. See Supplement 1 (Sec.~S3, Fig.~S3) for further details. 

To prepare each hydrophone, a single-sided tapered optical fibre was bonded to the paddle and coupled to the waveguide of the device at the optimal coupling position using a UV-curable adhesive (Norland Optical Adhesives 86H) and the device was lifted off the chip. The device bonded to the tapered fibre was then coated in two layers of PDMS (Sylgard 184, Sigma-Aldrich), which served to both facilitate the sensing mechanism and protect the 1D PhC cavity from damage. See Supplement 1 (Sec.~S3, Fig.~S4) for further details.

\subsection{Modelling Photonic Hydrophone Response}

Using COMSOL Multiphysics, the optical resonances of the photonic hydrophone and their changes in response to applied pressures were modelled. We considered the silicon PhC region of the hydrophone, encapsulated in a 1~{\textmu}m-thick PDMS coating. The optical resonances were determined using a three-dimensional eigenmode solver configuration. The electric field distribution of one such resonance is depicted in Fig.~\ref{Fig: modelling figure}(b). Here it is observed that the optical resonance is contained around the defect of the 1D PhC, with the evanescent field of the optical mode extending significantly into the surrounding media.

The expected shift in optical resonance frequency ($\delta \omega_0$) for an external pressure $ \delta\sigma$ is given by $G$, which describes the stress-optic coupling:
\begin{equation}
    G=\frac{\delta\omega_0}{\delta\sigma}.
    \label{Eq:G}
\end{equation} 

\noindent The value of $G$ is crucial for estimating the performance of the hydrophone, as it allows for the estimation of the shot-noise limited minimum detectable pressure (without considering mechanical contributions):

\begin{equation}
    \sigma^\textrm{SN}_\textrm{min} = \frac{1+R_d}{1-R_d} \frac{\omega_0}{Q_0G}\sqrt{\frac{\hbar\omega_\textrm{L}}{2\eta_\textrm{qe}P_\textrm{det}}} ,
    \label{EQ:SN}
\end{equation}

\noindent where $R_d$ is the reflectivity on the optical resonance, $\omega_0$ is the optical resonance frequency, $Q_0$ is the optical quality factor, $\omega_L$ is the laser frequency, $\eta_\textrm{qe}$ is the quantum efficiency of the photodetector and $P_\textrm{det}$ is the the detected optical power \cite{krause2012high,aspelmeyer2014cavity,schottky1918spontane,blanter2000shot}. See Supplement 1 (Sec.~S2 and Fig.~S2) for further details.

To model $G$ for the hydrophone, an optical eigenmode solver was performed while changing the optical properties of the hydrophone due to pressure. Within the frequency range of interest (0-200~kHz), the acoustic wavelength in water is many orders of magnitude greater than the largest dimension of the sensing region of the device, so the pressure of an acoustic wave is uniformly applied to the resonator. Further, the coatings considered (\textit{e.g.} PDMS) are optically transparent at 1550~nm and are acoustically impedance matched with water, which minimises acoustic reflections off the water-PDMS interface \cite{guillermic2019pdms}. Therefore we assume that around our sensing region the stress is uniformly distributed.

We modelled the response of the hydrophone to acoustic pressure through two different mechanisms - via photoelastic-mediated coupling and pressure-deformation optomechanical coupling. For the photoelastic-mediated coupling, we considered the contribution from the silicon device and the PDMS coating separately. By considering these three contributions, we determine which contribution is most dominant in the acoustic sensing mechanism of our hydrophone.


\label{subsec:Photoelastic}
The photoelastic effect, or stress-optic coupling, induces a change in refractive index $n$, which is proportional to the change in $\omega_0$, due to a stress $\sigma$ in a material \cite{nelson1971theory}. The stress-dependant refractive index $n(\sigma)$ can be written as:

\begin{equation}\label{Eq: photoelastic}
n(\sigma) = n_0+\frac{dn}{d\sigma}\sigma.
\end{equation}

\noindent where $n_0$ is the nominal refractive index, which is the refractive index of the material in the absence of acoustic pressure (when $\sigma=0$). For small strain,

\begin{equation}
\frac{dn}{d\sigma}\approx \frac{dn}{d\sigma}\Big|_{\sigma = 0},
\end{equation}

\noindent and the relation can be approximately linearized as:

\begin{equation}\label{Eq:photoelastic simplified}
n(\sigma) = n_0 + p_\text{mat}\sigma,
\end{equation}

\noindent where $p_\text{mat}$ is the materials photoelastic coefficient, which is proportional to $G$. We considered the photoelastic contribution from both the silicon device and the PDMS coating.

First considering silicon, which has a photoelastic coefficient of $p_{\textrm{Si}} = -2.463\times 10^{-11}$~Pa$^{-1}$ \cite{dickmann2023temperature} the stress-optic coupling ($G$) was determined by tracking the shift of the eigenfrequency of the optical mode with applied pressure, as depicted in Fig.~\ref{Fig: modelling figure}(c). The photoelastic contribution from the silicon device was found to be $G=125~\textrm{Hz/Pa}$. This value is very small, especially when considering shifts in the eigenfrequency at small pressures. However, as shown in Fig.~\ref{Fig: modelling figure}(b), a significant portion of the electric field is contained in the media surrounding the silicon device. Thus, the contribution from the coating layer must also be accounted for.

Assuming that the coating layer (\textit{e.g.} PDMS) is approximately homogeneous and isotropic, the change of index can be accounted for using the Clausius-Mossotti relation \cite{rysselberghe2002remarks}, which correlates the refractive index to the material density ($\rho$). This transforms to the Lorentz-Lorenz relation when $\epsilon_\textrm{r}=n^2$ \cite{giordano2000fiber}, where $\epsilon_\textrm{r}$ is the dielectric constant of the material, as:
 
 \begin{equation}\label{Eq: lorentz-lorenz} 
\frac{n^2 - 1}{n^2 + 2} = \frac{N}{3M\epsilon_0}\rho\beta,
\end{equation}

\noindent where $M$ and $\beta$ represent the molar mass and dipole polarizability of the material respectively, $N$ is Avogadro's number, and $\epsilon_0$ is the permittivity of free space \cite{park2018investigating}. The pressure dependant density of the material can be expressed as $\rho(\sigma) = \rho_0(1 - \sigma/E)$, where $E$ and $\rho_0$ represent the Young's modulus and nominal density at standard conditions. Rearranging Eq.~\ref{Eq: lorentz-lorenz} to directly relate $n$ and $\sigma$ yields the expression:

\begin{equation}\label{Eq: lorentz-lorenz rewrite}
n(\sigma) = \sqrt{\frac{3M\epsilon_0 + 2N\beta\rho_0(E - \sigma)}{3M\epsilon_0 - N\beta\rho_0(E - \sigma)}}.
\end{equation}

\noindent Additionally, it can be assumed that $\beta$ has negligible density dependence, and thus no pressure dependence (\textit{i.e.} $d\beta/d\sigma$ = 0) \cite{galbraith2014photoelastic}. The pressure-dependant refractive index change in the small strain regime can then be evaluated as:

\begin{equation}\label{Eq: lorentz-lorenz derivative, simplified}
    \frac{dn}{d\sigma} \approx \frac{9}{2}\frac{1}{E}\frac{MN\beta\rho_0\epsilon_0}{\sqrt{(3M\epsilon_0 - N\beta\rho_0)^3(3M\epsilon_0 + 2N\beta\rho_0)}}.
\end{equation}  

For the material properties of PDMS, we find that $dn/d\sigma\equiv p_\textrm{PDMS}\approx -9.08\times10^{-8}~\textrm{Pa}^{-1}$. This is approximately 3 orders of magnitude larger than $p_\textrm{Si}$, which implies that the shift in refractive index of PDMS will be much greater when compared to silicon for the same induced pressure. We confirm this by using COMSOL to determine the shift in eigenfrequency with pressure and extract the corresponding $G$ value, as presented in Fig.~\ref{Fig: modelling figure}(c). We find that $G=-2~\textrm{MHz/Pa}$, which is 6 orders of magnitude greater than that determined for the contribution from the silicon alone.

To determine the effect of pressure-induced structural deformations on the optical resonances, a uniform boundary stress, corresponding to the acoustic pressure, was applied to the hydrophone and the optical resonances were observed. This yielded $G=5.5~\textrm{kHz/Pa}$ (see Supplement 1 (Sec.~S1 and Fig.~S1)  for further details), which is 3 orders of magnitude smaller than that $G$ determined from the photoelastic-mediated contribution from the PDMS coating. Thus we conclude that the pressure-deformation coupling contribution to the sensing mechanism is also negligible, and that the photoelastic response of the PDMS coating is the dominant contribution to the sensing mechanism of our hydrophone.

The FEM modelling was also performed for several alternative coatings which were considered as candidates for the hydrophones, other than PDMS. We compared a low-refractive index protective coating (MY-133, MY Polymers) and a high viscosity version of the same coating (MY-133 V2000, MY Polymers). We found that the value of $G$ for PDMS was a factor of 2 greater than those of both low-refractive index coatings, which is primarily attributed to PDMS having a lower Young's modulus value. This indicated that PDMS is the best choice of coating out of those considered.



\section{Characterising Photonic Hydrophone Response}\label{Section: characterisation}

\subsection{Experiment Set-Up}\label{Section: characterisation set up}

\begin{figure}[htbp]
\centering\includegraphics[width=\textwidth]{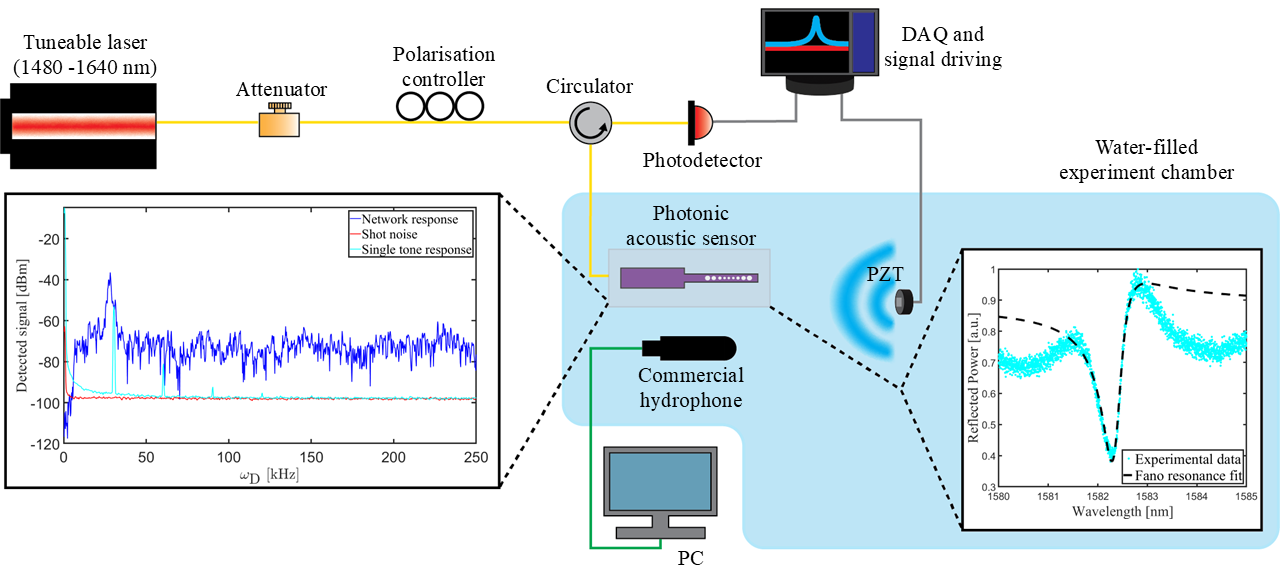}
\caption{\label{Fig: experimental set up with NR}Diagram of experiment set-up used for characterisation of photonic hydrophone. The blue-shaded region represents the water-filled experiment chamber that our hydrophone, the commercial hydrophone and PZT are submerged in. The network response from 0-250~kHz and a single tone response at 30~kHz from our hydrophone (see inset figure on left) and the optical reflection spectrum of our hydrophone while submerged underwater (see inset figure on right) are also displayed.}
\end{figure}

The response from our hydrophone was characterised using the set-up displayed in Fig.~\ref{Fig: experimental set up with NR}. A tunable laser (TSL-770, Santec) is swept from 1480-1640~nm and an optical mode of the device is identified. The optical modes are typically around 1550~nm, with an optical quality factor on the order of $\sim$$10^3$ in air. The optical mode shifts by $\sim$20~nm and the optical quality factor decreases by a factor of $\sim$10 once the device is submerged in water (see Fig.~\ref{Fig: experimental set up with NR}, and Supplement 1 (Sec.~S4, Fig.~S5) for further details). The experiments are typically conducted with an optical input power of 5~mW. To characterise the response of the hydrophone to an acoustic signal, both the hydrophone and a PZT are submerged and separated by 60~cm, in a water-filled experiment chamber. The experiment chamber is lined with an open-cell acoustic foam to reduce in-container reflections of acoustic signals. The PZT transduces a $2\,\text{V}_\text{pp}$ sinusoidal electrical signal into an acoustic signal, sweeping from 0-250~kHz, and a network response is taken with the hydrophone. A single-tone response is also taken at 30~kHz, and a network response is taken using a commercial hydrophone (Miniature Hydrophone 8103, Hottinger Brüel $\&$ Kjær). The commercial hydrophone network response is used to calibrate the amplitude of the PZT signal, which allows us to quantify the sensitivity of our hydrophone (see Section~\ref{Section: results}). All signal driving and data acquisition (DAQ) was conducted using a multi-test and measurement device (Moku:Pro, Liquid Instruments). 

\subsection{Characterisation Results}\label{Section: results}

\begin{figure}[htbp]
\centering\includegraphics[width=0.7\textwidth]{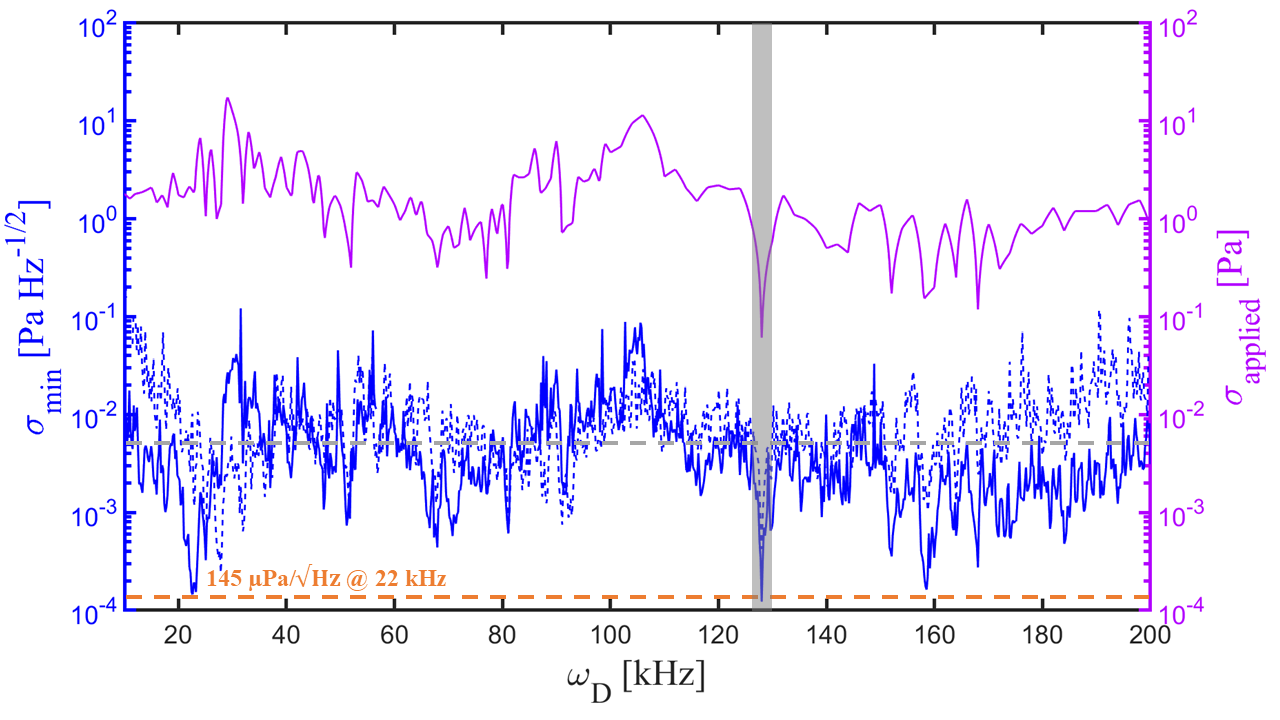}
\caption{\label{Fig: sensitivity} Sensitivity of our hydrophone, from 10-200~kHz, calibrated using commercial hydrophone. The blue-dashed line corresponds to the sensitivity calculated from another photonic hydrophone, to demonstrate the repeatability of the procedure (see Supplement 1 (Sec.~S5, Fig.~S6) for further details). Note that the grey-dashed line indicates the detection limit (minimum detectable pressure) of the commercial hydrophone. The purple line represents $\sigma_{\textrm{applied}}$, determined from the network response from the commercial hydrophone.}
\end{figure}

The measured response to acoustic signals is presented in Fig.~\ref{Fig: experimental set up with NR} and Fig.~\ref{Fig: sensitivity}. Fig.~\ref{Fig: experimental set up with NR} displays the response from our hydrophone when a $2\,\text{V}_\text{pp}$ signal at 30~kHz is driving with the PZT, where there is a corresponding sharp peak at 30~kHz. Fig.~\ref{Fig: experimental set up with NR} also displays the network response of our hydrophone. We note that our response is shot-noise limited across the majority of the frequency range. To calculate the sensitivity, as shown in Fig~\ref{Fig: sensitivity}, the network response and noise floor are used to calculate the signal-to-noise ratio (SNR). The SNR is then calibrated with the network response from the commercial hydrophone, which results in the sensitivity. The sensitivity can be defined as the minimum detectable pressure at some driven acoustic frequency $\omega_\textrm{D}$ \textit{i.e.}, SNR is equal to 1. The minimum detectable pressure can then be determined from \cite{basiri2019precision}:

\begin{equation}\label{Eq: sensitivity}
    \sigma_{\textrm{min}}(\omega_\textrm{D}) =  \frac{\sigma_{\textrm{applied}}(\omega_\textrm{D})}{\sqrt{SNR\times RBW}}.
\end{equation}

\noindent $\sigma_{\textrm{applied}}$ is the applied pressure, which is calibrated with the commercial hydrophone, and $RBW$ is the spectrum analyser resolution bandwidth. Note that the sensitivity is consistently on the order of $\sim$\SUm{1}, and reaches sub-\SUm{1} in several regions, with a peak sensitivity of 145~\SUu{1} (at 22~kHz). This experimentally observed sensitivity is consistent with that predicted by our modelling (see Eq.~\ref{EQ:SN}), $\sigma^\textrm{SN}_\textrm{min} = 4.1$~\SUm{1}, for an optical resonance with $\kappa = 0.5$~nm and $Q_0 = 3164.8$. The $\kappa$ and $Q_0$ values are extracted by fitting a Fano resonance \cite{navarathna2024silicon,limonov2017fano} to the experimental data, as described in Supplement 1 (Sec.~S2, Fig.~S2). Fig.~\ref{Fig: sensitivity} also displays the network response from the commercial hydrophone, equivalent to $\sigma_{\textrm{applied}}$. By comparing the plot of $\sigma_{\textrm{applied}}$ to $\sigma_{\textrm{min}}$, we can identify a sharp drop at $\sim$130~kHz (grey-shaded region). This dip in $\sigma_{\textrm{min}}$ is an unreliable feature of the plot, related to the dip in $\sigma_{\textrm{applied}}$ at the same frequency that is most likely a feature of the experiment chamber. Also note that there is improved sensitivity between 20-30~kHz. This is most likely a characteristic of the hydrophone construction, such as inconsistency in the thickness of the PDMS coating that encapsulates the silicon device. 

It should be noted that the commercial hydrophone cannot resolve signals below the order of a \SUm{1}, and so the region between the grey-dashed line in Fig.~\ref{Fig: sensitivity} identifies the region in which our hydrophone out-performs the commercial hydrophone. Further, the commercial hydrophone is limited to the frequency range 0-200~kHz. While we cannot calibrate the sensitivity of our hydrophone above 200~kHz, if we extend to network response beyond 200~kHz, we continue to observe a response from the our hydrophone. This implies that our hydrophone is operational across a larger frequency range than the commercial hydrophone. Furthermore, our hydrophone is 10 orders of magnitude smaller in volume than the commercial hydrophone. 

\section{Photonic Hydrophone Deployment}\label{Section: deployment}

\begin{figure}[htbp]
\centering\includegraphics[width=0.7\textwidth]{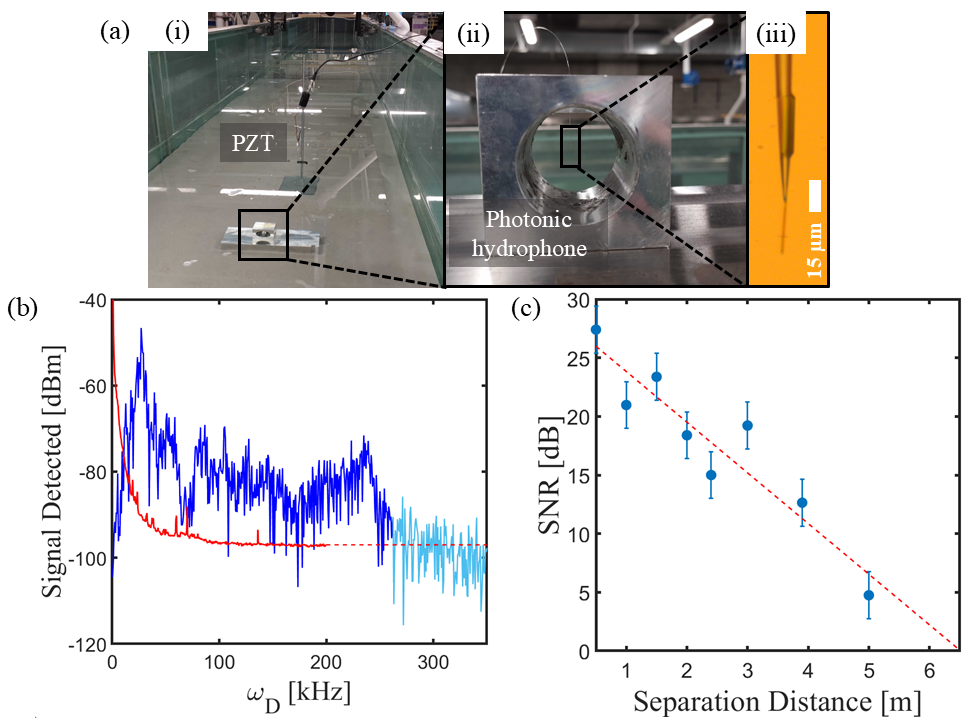}
\caption{\label{Fig: deployment} (a) Deployment of photonic hydrophone in a wave flume. (i) The photonic hydrophone and PZT submerged underwater in the wave flume. (ii) The photonic hydrophone mounted on a metal block. (iii) Microscope image of a prepared photonic hydrophone. (b) Network response of the photonic hydrophone. The blue line represents the network response, with the change to a light blue line indicating when the response sinks into the noise floor. The red line (dashed and solid) is the noise floor. (c) SNR of a 30~kHz single tone response with varying separation distance between the PZT and photonic hydrophone. The blue dots are the data points and the red dashed line corresponds to the linear trend-line. }
\end{figure}

Following characterisation of the photonic hydrophone, we deployed the hydrophone in a large wave flume, primarily used for engineering applications, to evaluate the suitability of the hydrophone for underwater measurement and communication. Specifically, we investigated the bandwidth and signal decay rate of the photonic hydrophone. Similar to the characterisation procedure, we submerged a PZT and the photonic hydrophone underwater (see Fig.~\ref{Fig: deployment}(a)), separated by 1.5~m.

To determine the bandwidth of our hydrophone, the PZT transduced a $10\,\text{V}_\text{pp}$ amplitude electrical signal into an acoustic signal, sweeping from 0-350~kHz, and a network response was taken with our hydrophone (see Fig.~\ref{Fig: deployment}(b)). By observing where the detected signal sinks into the noise floor, we determine that the bandwidth of our hydrophone is approximately 10-250~kHz. The poor SNR below 10~kHz is attributed to low frequency noise from engineering machinery in use during data collection. 

To determine the signal decay rate of the photonic hydrophone, we measured 30~kHz single tone responses for separation distances between the photonic hydrophone and PZT from 1.5~m to 5~m. Then by plotting the SNR of the single tone responses against separation distance and fitting a linear trend-line, we observe that the SNR decays with distance by approximately $4.3 \textrm{ dB m}^{-1}$ (see Fig.~\ref{Fig: deployment}(c)).

We further study the qualitative behaviour of the data by comparing the experimental data to a finite element simulation (see Supplement 1 (Sec.~S6)). In Fig.~\ref{Fig: deployment}(c), we observe that overall, the detected acoustic signal decays, and also oscillates due to interferences with reflected wave-fronts. The COMSOL simulation  demonstrates that due to an acoustic impedance mismatch on the water-glass and water-air interfaces, the acoustic wave reflects off these interfaces and interferes with itself. This results in an observed oscillating acoustic intensity with distance, which we also observe between individual data points in the experimental data (Fig.~\ref{Fig: deployment}(b)). From the simulation, we also observe that the acoustic signal decays overall, and that the main source of acoustic absorption is the sand at the bottom of the wave flume.
 
In an environment without boundaries, the acoustic intensity decays at a rate of $1/r^2$, where $r$ is the distance from the acoustic source \cite{bass2003physical}. In this regime, one would expect a signal decay of a factor of 11 when changing from $r = 1.5$~m to $r=5$~m. However, we observe a signal decay of a factor $\sim$100. Hence, the difference in the decay rate that we observe when conducting our experiment in a wave flume is likely due to the combination of refocusing of diffracted waves at boundaries and absorption from the sand as is observed at short length scales in the simulation (see Supplement 1 (Sec.~S6, Fig.~S7)).



\section{Conclusion}


We have developed a micron-sized free-standing 1D PhC-based silicon photonic hydrophone and have demonstrated that the hydrophone can detect $\sim$\SUm{1} acoustic signals across frequencies from 10-200~kHz, with a minimum detectable pressure of 145~\SUu{1} at 22~kHz. Our hydrophone demonstrates comparable sensitivity to a commercial hydrophone, while being far smaller, with the length of the sensing region being reduced from millimetres to tens of micrometres, and also reducing power requirements. We also deployed our hydrophone in a wave flume to evaluate its suitability for underwater measurement and communication, and demonstrate a communication bandwidth of 10-250~kHz and a signal decay rate of $4.3 \textrm{ dB m}^{-1}$. These advantages of our hydrophone over commercial hydrophones could enable high spatial resolution and non-invasive imaging of sub-micron sized acoustic features (\textit{e.g.}, living cell vibrations), and would also be advantageous for underwater communication and imaging applications. 



\begin{backmatter}

\bmsection{Funding} Next Generation Technologies Fund (NGTF) and Advanced Strategic Capabilities Accelerator (ASCA); Australian Research Council Centre of Excellence for Engineered Quantum systems (EQUS, Grant No. CE170100009); Australian Research Council Centre of Excellence in Quantum Biotechnology (QUBIC, Grant No. CE230100021); Air Force Office of Scientific Research (Grant No. FA9550-20-1-0391) 

\bmsection{Acknowledgments}The authors acknowledge the facilities, and the scientific and technical assistance, of the Australian Microscopy $\&$ Microanalysis Research Facility at the Centre for Microscopy and Microanalysis, The University of Queensland. This work was performed in part at the Queensland node of the Australian National Fabrication Facility, a company established under the National Collaborative Research Infrastructure Strategy to provide nano and microfabrication facilities for Australia’s researchers. The Commonwealth of Australia (represented by the Defence Science and Technology Group) supports this research through a Defence Science Partnerships agreement. This work was funded under the NGTF and is being delivered through the ASCA. The authors also thank Dr. Scott Foster (Defence Science and Technology Group) for his support, expertise and discussions on the research in this publication. This work was also  financially supported by the Australian Research Council Centre of Excellence for Engineered Quantum Systems (EQUS, Grant No. CE170100009), the Australian Research Council Centre of Excellence in Quantum Biotechnology (QUBIC, Grant No. CE230100021) and the Air Force Office of Scientific Research. The authors also thank the staff at the UQ School of Civil Engineering Hydraulics Laboratory for the use of their wave flume facilities. 

\bmsection{Disclosures}The authors declare no conflicts of interest.

\bmsection{Data availability}Data underlying the results presented in this paper are not publicly available at this time but may be obtained from the authors upon reasonable request.

\bmsection{Supplemental document}See Supplement 1 for supporting content.

\end{backmatter}

\bibliography{references}

\end{document}